\documentclass[12pt]{article}

\usepackage{graphicx}

\newcommand{\blackslug}{\hbox{\hskip 1pt
        \vrule width 4pt height 8pt depth 1.5pt\hskip 1pt}}
\newcommand{\myQED}{\hfill \blackslug}
\newcommand{\la}{\langle}
\newcommand{\ra}{\rangle}

\newenvironment{proof}
    {\pagebreak[1]{\narrower\noindent {\bf Proof:\nopagebreak}}}%
    {\myQED}

    {\myQED}

\newtheorem{lemma}{Lemma}

\begin{document}

\begin{center}
{\large \bf
Creating Teams of Simple Agents for Specified Tasks: \\
A Computational Complexity Perspective}

\vspace*{0.2in}

Todd Wareham \\
Department of Computer Science \\
Memorial University of Newfoundland \\
St.\ John's, NL Canada \\
(Email: {\tt harold@mun.ca}) \\

\vspace*{0.1in}

\today
\end{center}

\begin{quote}
{\bf Abstract}:
Teams of interacting and co-operating agents have been proposed as
an efficient and robust alternative to monolithic centralized control for
carrying out specified tasks in a variety of applications. A number of
different team and agent architectures have been investigated, e.g.,
teams based on single vs multiple behaviorally-distinct types of
agents (homogeneous vs heterogeneous teams), simple vs complex agents, direct
vs indirect agent-to-agent communication. 
A consensus is emerging that (1) heterogeneous teams composed
of simple agents that communicate indirectly are preferable 
and (2) automated methods for verifying and designing 
such teams are necessary.
In this paper, we use computational complexity analysis to assess viable algorithmic
options for such automated methods for various types of
teams. Building on recent complexity analyses addressing related questions in swarm 
robotics, we prove that automated
team verification and design are by large both exact and approximate
polynomial-time intractable in general 
for the most basic types of homogeneous and heterogeneous teams
consisting of simple agents that communicate indirectly. Our results
suggest that tractability for these problems must be sought relative to additional 
restrictions on teams, agents, operating environments, and tasks.
\end{quote}

\section{Introduction}

Teams of interacting and co-operating agents have been proposed as
an efficient and robust alternative to monolithic centralized control for
carrying out specified tasks in a variety of applications. A number of
different team and agent architectures have been investigated. Three
dimensions of these architectures are of particular importance:

\begin{enumerate}
\item Should teams consist of a single type of agent or multiple
       types of agents, i.e., should teams be homogeneous or heterogeneous?
\item Should individual agents have simple or complex control mechanisms, i.e.,
       simple reflex or model- / goal- / utility-based agents 
       \cite[Section 2.4]{RN10}?
\item Should individual agents communicate directly with each other by
       agent-to-agent messages or indirectly via their sensed presences and/or 
       environmental modifications, i.e., via stigmergy \cite{BDT99}?
\end{enumerate}

\noindent
Based on the experience gained with various proof-of-concept experiments and
implementations, the consensus is emerging that (1) heterogeneous teams composed
of relatively simple agents that communicate indirectly are preferable 
\cite{DF+13,KH+15,SS+21} and (2) automated methods for verifying and designing 
such teams are necessary \cite{BB+17,DTT20}. 

A natural question at this point is what algorithmic options are and are not 
available for the efficient verification and design of teams relative to the 
three dimensions listed above.  In this paper, we give some initial answers
to this question, building on recent work \cite{TW19,War19,WdH+Sub,WV18_SI,WV18_AT,WV21}
addressing related questions in swarm robotics for distributed construction.\footnote{
Though there has been other complexity-theoretic work on individual agent verification 
and design \cite{DLW03,Ste03,WD02}, the models of agents and environments used were too 
abstract to allow examination of agent teams, simple reflex agents, or stigmergy.
}
In particular, we give proofs (modified from several given
previously in \cite{War19,WdH+Sub,WV18_SI}\footnote{
These modifications are described along with the proofs of our results in the
appendix.
}) which demonstrate that the problems
of team verification and design are by large both exact and approximate
polynomial-time intractable in general 
relative to the most basic types of homogeneous and heterogeneous teams
consisting of simple reflex agents that do not use stigmergy. This in turn
suggests that tractability must be sought relative to additional restrictions
on teams, agents, operating environments, and tasks.

\section{Methods}

In this section, we first review the basic entities in our model of task 
performance by robot teams --- namely, environments, individual robots, robot 
teams, and tasks (with the last of these being new to this paper). Though 
this is a basic model in which robots sense and
move without uncertainty in a discrete and synchronous manner in a 2D
grid-based environment, it is flexible enough to allow investigations along\
the three team and agent architectural dimensions listed in the introduction.
In the interests of concision, most of this review is the ``short form'' given 
in \cite{TW19,WdH+Sub}; readers wishing more details should consult \cite{War19,WV18_SI}.
This will be followed by formalization of computational 
problems within this model corresponding to various types of robot team 
verification and design.

The basic entities in our model are as follows:

\begin{itemize}
\item{
{\bf Environments}: Our robots operate in a finite 2D
square-based environment $E$ in which each square is either a freespace (which
a robot can occupy or move through) or an obstacle (which a robot cannot 
occupy or move through), and has a square-type, e.g.,
grass, gravel, wall, drawn from a set $E_T$. Let $E_{i,j}$ denote the square 
that is in the $i$th column and $j$th row of $E$ such that $E_{1,1}$ is the 
square in the southwest-most corner of $E$.
}
\item{
{\bf Robots}: Each robot occupies a square in $E$ and in a 
basic movement-action can either move exactly one square to the north, south,
east or west of its current position or elect to stay at its current position.
Each robot has a sensing-distance bound $r$ such that the robot can sense the type 
of the square at any position within Manhattan distance $r \geq 0$ of the robot's 
current position (with $r = 0$ corresponding to the square
on which the robot is standing). These square-types are accessible via
predicates of the form $enval(e,pos)$ which returns $True$ if the square at position
$pos$ has type $e \in E_T \cup \{e_{robot}\}$ (with the sensor returning $e_{robot}$
if a robot is occupying square $pos$) and $False$ otherwise, where a position
$pos$ is specified in terms of a pair $(x,y)$ specifying an environment-square
$E_{i+x,j+y}$ if the robot is currently occupying $E_{i,j}$. Each robot can change the
type of the square at any position within Manhattan distance one of the robot's
current position to type $e$ via predicates of the form $enmod(e,pos)$ where $pos$ is
specified as for $enval()$. 

Each robot has a finite-state controller and is hence known as a Finite-State Robot 
(FSR). Each such controller consists of a set $Q$ of states linked
by transitions, where each transition $(q, f, x, dir, q')$ between states 
$q$ and $q'$ has a propositional logic trigger-formula $f$, an environment 
modification specification $x$, and a movement-direction $dir \in \{goNorth, goSouth, 
goEast, goWest, stay\}$. Trigger-formulas and 
modification specifications are typically stated in terms of predicates $enval()$ and 
$enmod()$, respectively. Both of these
specifications can also be stated as a special symbol $*$, which is interpreted as 
follows: If $f \neq x \neq *$ and the transition's trigger-formula evaluates to $True$, 
i.e., the transition is enabled, this causes the environment-modification specified by 
$x$ to occur, the robot to move one square in direction $dir$, and the robot's state
to change from $q$ to $q'$. If $f = *$, the transition executes if no other transition 
executes (making this in effect the default transition); if $x = *$, no 
environment-modification is made. 

The transitions in the FSR described above can be viewed as condition-action rules
within the agent framework given in \cite[Section 2.4]{RN10} such that the $*$ 
trigger formula on a transition leaving state $q$ can be viewed as the negation of the 
disjunction of the trigger formulas of all other transitions leaving $q$. Given this,
single- and multi-state FSR correspond to simple reflex and model-based reflex
agents agents, where the model and UPDATE-STATE() function in \cite[Section 2.4.3]{RN10}
are implicit in the transitions between states. Note that actions correspond to
FSR movements and environmental modifications as required, and are flexible enough
to allow situations in which FSR movements or environment modifications do not occur.
}
\item{
{\bf Robot teams}: A team $T$ consists of a set of the robots described above, where
there may be more than one robot with the same controller on a team; as such, we allow
both homogeneous and heterogeneous teams. Let $T_i$ denote the $i$th robot on the team. 
Each square in $E$ can hold at most one member of $T$; if at any point in the execution of
a task two robots in a team attempt to occupy or modify the same free space or a robot 
attempts to occupy the same space as an obstacle, the execution terminates and is 
considered unsuccessful. A {\bf positioning} of $T$ in $E$ is an assignment of the robots 
in $T$ to a set of $|T|$ squares in $E$. For simplicity, team members move synchronously, 
and once movement is triggered, it is atomic in the sense that the specified movement is
completed. 

Note that robots in our teams do not communicate with each
other directly --- rather, they can communicate with each other indirectly 
through their sensed presences in and changes they make to the environment,
i.e., via stigmergy \cite{BDT99}. In the remainder of the paper, we will find
it useful to distinguish  these two types of communication, which
will be denoted as {\bf agent-} and {\bf environment-based stigmergy}, respectively. 
}
\item{
{\bf Tasks}: Tasks are specified in terms of a desired set of environment square-values,
robot positions, and/or robot internal states, e.g., a $3 \times 3$ square of
square-type $e_X$ has been created at a particular location in the environment,
all robots in a team are in state $q_F$ and located on the eastmost edge of the 
environment. Such is a specification will be denoted as a task's {\bf target 
configuration}. We will assume that for each task $Tsk$ and an environment $E$ in which
a robot team $T$ is operating, $E$ can be checked for the target 
configuration associated with $Tsk$ in time polynomial in the size of $E$.
}
\end{itemize}

\noindent
We use the notions of deterministic robot and team operation 
proposed in \cite{TW19,WV18_SI} (i.e., requiring that at any time as the team 
operates in an environment, all transitions enabled in a robot relative to
the current state of that robot perform the same environment modifications
and progress to the same next state). Given this, an individual FSR is not itself 
deterministic but rather the operation of that FSR is deterministic in the context of a
particular FSR team operating in a particular environment.

Let us now consider the team verification design problems that we will analyze in
this paper, starting with verification.

\vspace*{0.06in}

\noindent
{\sc Team / environment verification for Task $Tsk$} (TeamEnvVer) \\
{\em Input}: An environment $E$ based on square-type set $E_T$, a an FSR team $T$,
              an initial positioning $p_I$ of $T$ in $E$, and an integer 
              $\#_{ec} \geq 0$. \\
{\em Question}: Does $T$ started at $p_I$ perform task $Tsk$ while making at most
                 $\#_{ec}$ square-type changes in $E$?

\vspace*{0.06in}

\noindent
We will consider two types of robot team design. Both types of design are done 
relative to a given design library. In the first case, $L$ consists of
transition templates of the form $(q, f, x, move, q')$ which are used 
to construct FSR controllers from a specified set of states by instantiating transition 
templates relative to those states. Note it may be the case that $q = q'$ in such a 
construction, i.e., a transition may loop back on the same state.

\vspace*{0.06in}

\noindent
{\sc Controller design by library selection for task $Tsk$} (ContDesLS) \\
{\em Input}: An environment $E$ based on square-type set $E_T$, a requested team-size 
              $|T|$, an initial positioning $p_I$ of $T$ in $E$, 
              a transition template library $L$, and
              integers $r \geq 0$, $|Q| \geq 1$, $d \geq 1$, $h \geq 1$, and
              $\#_{ec} \geq 0$. \\
{\em Output}: A controller-set $C$ in which each controller has sensory radius $r$, 
               at most $|Q|$ states, and at most $d$ transitions chosen from $L$ out of 
               any state such that an FSR team $T$ based on $h$ controllers from $C$
               started at $p_I$ performs $Tsk$ while making at most $\#_{ec}$ square-type
               changes in $E$, if such a $C$ and $T$ exists, and special symbol $\bot$ 
               otherwise.

\vspace*{0.06in}

\noindent
In the second case, $L$ consists of complete FSR controllers.

\vspace*{0.06in}

\noindent
{\sc Team design by library selection for task $Tsk$} (TeamDesLS)  \\
{\em Input}: An environment $E$ based on square-type set $E_T$, a requested
              team-size $|T|$, an FSR library $L$, an initial region $E_I$
              of size $T$ in $E$, and integers $h \geq 1$ and $\#_{ec} \geq 0$. \\
              $E$. \\
{\em Output}: An FSR team $T$ based on $h$ robots selected from $L$ such that $T$ started 
               in $E_I$ performs $Tsk$ while making at most $\#_{ec}$ square-type
               changes in $E$, if such a a $T$ exists, and special symbol $\bot$ 
               otherwise.

\vspace*{0.06in}

\noindent
We will subsequently analyze these problems relative to three parameters:

\begin{enumerate}
\item The number of different types of FSR controllers in a team ($h$);
\item The maximum number of states in the robots in a team ($|Q|$); and
\item The maximum allowable number of environmental changes made by a team in 
       performing specified tasks ($\#_{ec}$).
\end{enumerate}

\noindent
Different values of these parameters will allow us to investigate the
question posed in the introduction --- namely, the effects on the computational
difficulty of team verification and design when using
(1) teams based on single ($h = 1$) and multiple ($h > 1$)
types of FSR controllers (i.e., homogeneous and heterogeneous teams),
(2) simple reflex ($|Q| = 1$) and model-based reflex ($|Q| > 1$) agents, and 
(3) agents that do ($\#_{ec} > 0$) and do not ($\#_{ec} = 0$) use environment-based 
stigmergy in the performance of tasks.

\section{Results}

In this section, we analyze the computational difficulty of our team verification\
and design problems relative to the three restrictions proposed at the end of
the previous section. We evaluate this difficulty relative to several criteria
of efficient algorithm operation using two standard techniques -- namely, proving 
tractability by giving algorithms and intractability by giving reductions from known 
intractable problems (see \cite{AB09,GJ79} for details of these techniques).
All proofs of results are relegated to the appendix.

Let us first consider exact polynomial-time solvability. An exact polynomial-time 
algorithm is an algorithm which always produces the correct output for a given input 
and whose runtime is asymptotically upper-bounded, i.e., upper-bounded when $|x|$ 
goes to infinity, by $c|x|^{c'}$, where $|x|$ is
the size of the input $x$ and $c$ and $c'$ are constants. A problem that has
a polynomial-time algorithm is said to be {\bf polynomial-time tractable}. 
Polynomial-time tractability is desirable because runtimes increase slowly as 
input size increases, and hence allow the solution of larger inputs.
We start with team verification.\footnote{
Results A, B, D, and E hold relative to some combination of the conjectures 
$P \neq NP$ and $P = BPP$, which though unproven
are widely believed to be true within computer science \cite{For09,Wig07}.
}

\vspace*{0.06in}

\noindent
{\bf Result A} (Modified from Lemma 4 in the Supplementary Materials of \cite{WV18_SI}): 
                TeamEnvVer is not exact polynomial-time tractable when $|Q| = 1$, 
                 $h = 1$, and $\#_{ec} = 0$. Moreover, this
                 intractability holds for any version of TeamEnvVer when
                 $h > 1$, $|Q| > 1$, and $\#_{ec} > 0$.  
\vspace*{0.06in}

\noindent
This result demonstrates that verification is polynomial-time intractable in general for
homogeneous teams of simple reflex agents that do not use environment-based
stigmergy. This in turn motivates the notion (introduced in \cite{WV18_SI})
of \linebreak {\bf $(c_1,c_2)$-completability}, which requires that each robot team 
complete its task within $c_1(|E| + |Q|)^{c_2}$ timesteps for constants $c_1$ and 
$c_2$.\footnote{
For technical reasons that are described in detail in \cite{WdH+Sub}, 
$c_1$ and $c_2$ are not part of the problem input but are specified beforehand
To ensure generous but still low-order polynomial team runtime bounds, we will assume 
that $c_1 = 10$ and $c_2 = 3$.
}
Let the versions of ContDesLS and TeamDesLS with this completability restriction be
denoted by ContDesLS$_{res}$ and TeamDesLS$_{res}$. It turns out that this
restriction does not always help in general.

 \vspace*{0.06in}

\noindent
{\bf Result B} (Modified from Lemma 5 in \cite{WV18_SI}):
                ContDesLS$_{res}$ is not exact polynomial-time tractable when $|Q| = 1$, 
                 $h = 1$, and $\#_{ec} = 0$. Moreover, this
                 intractability holds for any version of ContDesLS$_{res}$ when
                 $h > 1$, $|Q| > 1$, and $\#_{ec} > 0$.  

\vspace*{0.06in}

\noindent
{\bf Result C} (Modified from Result A in \cite{War19}): 
                TeamDesLS$_{res}$ is exact polynomial-time tractable when $|Q| \geq 1$, 
                 $h = 1$, and $\#_{ec} \geq 0$.

\vspace*{0.06in}

\noindent
{\bf Result D} (Modified from Lemma A.7 in \cite{WdH+Sub}): 
                TeamDesLS$_{res}$ is not exact polynomial-time tractable when $|Q| = 1$, 
                 $h = 2$, and $\#_{ec} = 0$. Moreover, this
                 intractability holds for any version of TeamDesLS$_{res}$ when
                 $h > 2$, $|Q| \geq  1$, and $\#_{ec} \geq 0$.  

\vspace*{0.06in}

\noindent
The above demonstrates that (1) restricted controller design (like verification) is 
polynomial-time intractable in general for homogeneous teams of simple reflex 
agents that do
not use environment-based stigmergy (Result B) and (2) though restricted team design is 
polynomial-time tractable for {\em any} type of homogeneous team (Result C), it is 
polynomial-time intractable in general for the simplest heterogeneous teams based on 
simple reflex agents that do not use environment-based stigmergy (Result D). 

Let us now consider polynomial-time approximate solvability. This type of 
solvability may be acceptable in situations where always getting the correct output for 
an input is not required.  Three popular types of polynomial-time approximation 
algorithms are:

\begin{enumerate}
\item
algorithms that always run in polynomial time but are frequently correct in that they 
produce the correct output for a given input in all but a small number of 
cases (i.e., the number of errors for input size $n$ is bounded by function $err(n)$) 
\cite{HW12};
\item
algorithms that always run in polynomial time but are frequently correct in that they 
produce the correct output for a given input 
with high probability \cite{MR10}; and
\item
algorithms that run in polynomial time with high probability but are always correct
\cite{Gil77}.
\end{enumerate}

\noindent
Algorithms of type (2) are of particular interest as they include evolutionary 
algorithms. Unfortunately, none of these options are in general open to us either 
courtesy of the following result.

\vspace*{0.06in}

\noindent
{\bf Result E} (Modified from Results A.4 and A.5 in \cite{WdH+Sub}): 
                None of the intractable versions of TeamEnvVer, ContDesLS$_{res}$,
                 or TeamDesLS$_{res}$ described in Results A, B, and D are
                 polynomial-time approximable in senses (1--3).

\vspace*{0.06in}

\noindent
Note that all intractability results above hold relative to the simplest types of tasks
(do some subset of the robots in team $T$ reach particular positions in $E$?) and
(in the case of ContDesLS$_{res}$ and TeamDesLS$_{res}$) the most restrictive type of 
completability, i.e., $(1,1)$-completability.

\section{Discussion}

In the previous section, we demonstrated that agent team verification and
two types of design by library selection (agent controller and team) are
by large both exact (Results A, B, and D) and approximate (Result E) polynomial-time 
intractable in general relative to 
the simplest possible types of teams and agents, i.e., homogeneous teams consisting
of simple reflex agents that do not use environment-based stigmergy. Even in the one
case where we have polynomial-time tractability (Result C; homogeneous team design by 
library selection), intractability asserts itself when as few as two types of agents 
co-exist on a team (Result D).\footnote{
It is tempting to think that 
polynomial-time intractability of team verification and design follows from the 
well-known combinatorial explosion in the number of possible team states and
designs in multi-agent systems. However, there are many examples of
problems with such exponential-size search spaces that are nonetheless solvable
in polynomial time by algorithms that exploit structure in those spaces,
e.g., {\sc Minimum spanning tree} \cite[Chapter 23]{CLR+09}. Hence, definitive proof of
polynomial-time intractability requires proofs such as those given here.
}
That the various intractabilities we have demonstrated
cannot be vanquished by invoking verification and design relative to teams with
higher values of $h$, $|Q|$, and $\#_{ec}$ suggests that we are seeing
a tight frontier of tractability \cite[Section 4.1]{GJ79} relative to 
the lowest possible values of these parameters.\footnote{
It is also intriguing that agent-based stigmergy is critical to some (Results A and
D) but not all (Result B) of our intractability results. It is all too often assumed
in discussions about stigmergy that environmental modifications
(including ``smart'' materials \cite{WN06}) are key. Our proofs suggest that in certain
situations, environmental modifications are unnecessary given sufficiently large
and appropriately-structured groups of mobile agents. As large groups of agents 
are of increasing interest in certain applications, this warrants further investigation.
}

As all of our results are derived relative to a simplified team operation model in
which deterministic agents operate in a synchronous and discrete manner within a 2D 
grid-based environment, these results are not immediately applicable to probabilistic
agents that operate in an asynchronous and continuous manner in the real world.\footnote{
Results like ours do nonetheless have a surprising generality; the interested reader
is referred to Sections 5 and 6.2 of \cite{War19} and
Section 5 of \cite{WV18_SI} for details.
}
That being said, our results do for now offer some reasons for real-world roboticists to 
be cautious. In particular, the fact that team verification and design are 
polynomial-time intractable even when
agent motion and sensing are error-free and occur in completely-observable
environments hints that there may be additional 
sources of computational difficulty in these problems that are not associated with agent
motion and sensing under partial observability and uncertainty \cite{HW+21}. These
sources should be acknowledged and investigated, particularly if team verification and
design must behave both efficiently and correctly when fully automated without
human oversight.

This last point highlights a crucial caveat when interpreting our intractability
results -- namely, these results hold relative to a simplified model of agent team
operation and, perhaps more importantly, restrictions 
on the values of $h$, $|Q|$, and $\#_{ec}$. Our frequent proviso that intractability
results apply ``in general'' was not mere rhetoric --- tractability may well hold for the
cases we considered when additional restrictions are in place; then again, it may not. In
either case, this must be determined by future complexity analyses. Hence, our results 
should be seen not as final statements on the intractability of team verification and
design but rather as interim guidelines suggesting where in the universe of restriction
possibilities tractability does and does not hold.

Given the above, future research into team verification and design should perhaps
more closely incorporate computational complexity analyses like those given here.
Such research could initially focus on more fully characterizing those combinations
of restrictions that do and do not render team verification and design tractable.
Such work has already been started \cite{TW19,War19,WdH+Sub,WV18_AT,WV18_SI,WV21}
for team verification and design in swarm robotics relative a variety of restrictions 
using more advanced analysis techniques (e.g., parameterized
complexity analysis \cite{DF13}). Additional restrictions of particular interest here
are those that ``break'' the reductions underlying our intractability results, e.g.,
restrictions on the degree and type of structure encountered by agents in their
environments (including the presences of other agents). Once these initial intractability
maps have been derived for simplified team operation models, they should be extended
to more realistic models incorporating stochasticity and uncertainty. Part of this
can be done by using complexity analysis techniques that explicitly incorporate 
stochasticity
\cite{MR10,Don19,MM13}. Complexity-based frameworks that incrementally build on
simplified operation models in a systematic and principled manner to create results
applicable to more realistic models (analogous to those developed in linguistics 
\cite{Ris93} and cognitive science \cite{vRB+19}) may also be of use in this
endeavour.

\section*{Acknowledgments}
Funding for this work was provided by a National Science and Engineering
Research Council (NSERC) Discovery Grant to TW (grant \# 228105-2015).


\newpage

\appendix

\section{Proofs of Results}

\label{SectProof}

All of our intractability results will be derived using polynomial-time reductions
from the following problems:

\vspace*{0.1in}

\noindent
{\sc 3-Satisfiability} (3SAT) \cite[Problem LO2]{GJ79} \\
{\em Input}: A set $U$ of variables, a set $C$ of disjunctive clauses over $U$ 
              such that each clause $c \in C$ has $|c| = 3$. \\
{\em Question}: Is there a satisfying truth assignment for $C$?

\vspace*{0.1in}

\noindent
{\sc Dominating set} \cite[Problem GT2]{GJ79} \\
{\em Input}: An undirected graph $G = (V, E)$ and a positive integer $k$. \\
{\em Question}: Does $G$ contain a dominating set of size $k$, 
              {\em i.e.}, is there a subset $V' \subseteq V$, $|V'| = k$, such 
              that for all $v \in V$, either $v \in V'$ or there is at least one
              $v' \in V'$ such that $(v, v') \in E$?

\vspace*{0.1in}

\noindent 
For each vertex $v \in V$, let the complete neighbourhood $N_C(v)$ of $v$ be the
set composed of $v$ and the set of all vertices in $G$ that are adjacent to $v$ 
by a single edge, {\em i.e.}, $v \cup \{ u ~ | ~ u ~ \in V ~ \rm{and} ~ (u,v) 
\in E\}$. We assume below for each instance of {\sc Dominating set}
an arbitrary ordering on the vertices of $V$ such that $V = \{v_1, v_2, \ldots, 
v_{|V|}\}$; we also assume analogous orderings for the variables and clauses of each
instance of {\sc 3-Satisfiability} such that $U = \{u_1, u_2, \ldots, u_{|U|}\}$ and
$C = \{c_1, c_2, \ldots, c_{|C|}\}$.

For technical reasons \cite{AB09,GJ79}, our intractability results are initially
derived relative to decision versions of our problems, i.e., problems whose answers
are either ``Yes'' or ``No''. Problem TeamEnvVer is already phrased as a decision
problem. The decision versions of ContDes:S$_{res}$ and TeamDesLS$_{res}$ (denoted
by ContDesLS$_{res,D}$ and TeamDesLS$_{res,D}$, respectively) ask if the structures
requested in each problem ($(C,T)$ and $T$, respectively) exist or not. 
The following lemma (based on the observation that any algorithm for non-decision problem
{\bf X} can be used to solve {\bf X}${}_D$) will be useful
below in transferring results from decision problems to their associated
non-decision problems.

\begin{lemma}
If {\bf X}$_D$ is not solvable in polynomial time relative to conjecture \textbf{C} then
{\bf X} is not solvable in polynomial time relative to conjecture \textbf{C}.
\label{LemAppProp1}
\end{lemma}

All proofs of results given here are modifications of proofs given previously in 
\cite{War19,WdH+Sub,WV18_SI}. It is thus appropriate to
describe the nature of these modifications, starting with an overview. All 
previous proofs had 
$|Q| = 1$, so modifications were made to ensure 

\begin{enumerate}
\item the initial values of $h$ and $\#_{ec}$ stated in the first part of
       each result, and 
\item the higher values of $|Q|$, $h$, and $\#_{ec}$ stated in the second 
       part of each result.
\end{enumerate}

\noindent
 As previous work was done with respect to distributed 
constructions tasks, $\#_{ec} = 0$ was easy to ensure in the modifications 
once we introduced our task model (which was new to the current paper) and
focused on tasks that involve achieving specified environment / agent / state 
configurations --- we just needed robots to get to specified  environmental
positions where we previously had them also building structures in those 
positions. Other result-specific modifications are summarized below in
text prior to each result's proof.

We now present the proofs of results in our paper, starting with Result A. 
Part (1) of this result  requires relatively straightforward modification of
environments and FSR transitions to convert the previous $h = 5$ 
construction in \cite{WV18_SI} to the needed $h = 1$ construction while 
preserving $|Q| = 1$. Part (2) requires relatively straightforward 
modification to allow $|Q| > 1$ and $h > 1$. 

\vspace*{0.1in}

\noindent
{\bf Result A} (Modified from Lemma 4 in the Supplementary Materials of \cite{WV18_SI}): 
                TeamEnvVer is not exact polynomial-time tractable when $|Q| = 1$, 
                 $h = 1$, and $\#_{ec} = 0$ unless $P = NP$. Moreover, this
                 intractability holds for any version of TeamEnvVer when
                 $h > 1$, $|Q| > 1$, and $\#_{ec} > 0$ for fixed values of
                 $h$, $|Q|$, and $\#_{ec}$.  

\vspace*{0.1in}

\begin{proof}
Lemma 4 in \cite{WV18_SI} gives a reduction from 3SAT to problem ContEnvVer$_{syn}$,
which is our problem TeamEnvVer without the restriction on $\#_{ec}$. In the
instance of ContEnvVer$_{syn}$ created by this reduction, all possible truth
assignments to the variables in $U$ in the given instance of 3SAT are generated one at
a time by the movements of a team of $|U| + 3$ single-state FSR of 4 types ({\em Variable,
Carry, CarryDetect}, and {\em CarrySignal}) in an environment $E$ such that the values in
the truth assignment are encoded in the positions of the {\em Variable} FSRs. Each of 
these assignments is in turn checked by a single-state FSR of a fifth type ({\em 
Evaluate}) against the collection of clauses $C$ in the given instance of 3SAT, and if the
truth assignment satisfies the conjunction of the clauses in $C$, the {\em Evaluate} FSR 
moves one square to the east and modifies the type of the square it is now placed on.

We modify this reduction to create a reduction from 3SAT to TeamEnvVer when $h = 1$
and $\#_{ec} = 0$ as follows:

\begin{enumerate}
\item As each of the five types of FSR can only occupy very specific non-overlapping areas
       in $E$, change the types of the squares in the areas occupied by each type of FSR 
       to new FSR-type-specific square-types $e_{Var}$, $e_{Car}$, $e_{CD}$, $e_{CS}$, and
       $e_{Evl}$.
\item Create a single-state universal FSR that can simulate all five types of FSR by 
       combining modified versions of all transitions in the five types of FSR, where
       a transition of the form $\langle, q, f, m, q\rangle$ in FSR-type $t$ is changed to
       $\langle q, enval(e_t,(0,0)) \wedge f, m, q\rangle$, i.e., FSR-type-specific
       transitions can only trigger if the FSR is in the environmental area associated
       with FSR of that type.
\item Replace the environmental modification made by the {\em Evaluate} FSR with $*$.
\end{enumerate}

\noindent
Note that this reduction creates an FSR team in which $h = 1$, $|Q| = 1$, and
$\#_{ec} = 0$. If we then make the target configuration of the task associated
with this instance of TeamEnvVer be that an FSR is positioned immediately to the
west of the initial position of the {\em Evaluate} FSR, the proof of reduction
correctness in Lemma 4 also establishes that this reduction is correct. As 
3SAT is $NP$-hard \cite[Problem L02]{GJ79}, this reduction establishes that TeamEnvVer 
is also $NP$-hard when $h = 1$, $|Q| = 1$, and $\#_{ec} = 0$ and and hence not 
polynomial-time solvable under these restrictions unless $P = NP$.

We now need to establish the $NP$-hardness of TeamEnvVer for fixed values of 
$h > 1$, $|Q| > 1$, and $\#_{ec} > 0$. Observe that the instance of TeamEnvVer 
constructed above makes no environmental modifications and hence trivially makes at most 
$\#_{ec}$ environmental modifications for any fixed value of $\#_{ec}$. In the case of 
$h$ and $|Q|$, construct a modified instance of TeamEnvVer above in which there is 
an additional ``holding area'' in $E$ consisting of $h - 1$ squares enclosed by 
obstacle-squares such that no FSR inside this area can leave it. Populate this holding 
area with arbitrary FSR based on $|Q|$ states such that none of these FSR makes an 
environmental modification. As $h$ and $|Q|$ are of fixed value, this modified instance 
of TeamEnvVer can still be constructed in time polynomial in the size of the given 
instance of 3SAT. Moreover, the reduction from 3SAT to this modified instance of
TeamEnvVer shows the $NP$-hardness of TeamEnvVer for the specified values
of $h$ and $|Q|$.
\end{proof}

\vspace*{0.1in}

Part (1) of Result B requires relatively straightforward modification 
to use transition-template library $L$, which actually ends up simplifying
the original proof in \cite{WV18_SI}. As the team consisted of a single 
FSR, this trivially gives $h = 1$. Part (2) requires complex and decidedly 
non-trivial
modification to allow $|Q| > 1$ and relatively straightforward modification
of the construction used in the proof of part (2) of Result A to allow $h > 1$.

\vspace*{0.1in}

\noindent
{\bf Result B} (Modified from Lemma 5 in \cite{WV18_SI}):
                ContDesLS$_{res}$ is not exact polynomial-time tractable unless
                 $P = NP$when $|Q| = 1$, $h = 1$, and $\#_{ec} = 0$. Moreover, this
                 intractability holds for any version of ContDesLS$_{res}$ when
                 $h > 1$, $|Q| > 1$, and $\#_{ec} > 0$ for any fixed values of
                 $h$, $|Q|$, and $\#_{ec}$..  
 
\vspace*{0.1in}

\begin{proof}
Lemma 5 in \cite{WV18_SI} gives a reduction (based on a reduction in
\cite{WV18_AT}) from {\sc Dominating set} to
a problem ContDes$^{fast}_{D,syn}$ which is essentially ContDesLS$_{res,D}$
in which selection from a library $L$ of transition-templates is simulated
by specifying bounds in the problem input on $|f|$, the maximum length
of any transition trigger-formula. This reduction creates a somewhat complex 
environment for a team composed of a single-state FSR \cite[Figure 2(b)]{WV18_AT}.
In order to force the transitions in such a robot to encode 
a candidate dominating set of size $k$ in the graph $G$ in the given instance
of {\sc Dominating set}, the robot has 
to navigate from the southwestmost corner of the environment to the top of the 
$(k+1)$st column in subgrid SG1 \cite[Figure 2(c)]{WV18_AT}. From there, the 
robot navigates the $|V|$ columns of subgrid SG2 \cite[Figure 2(d)]{WV18_AT},
where each column represents the vertex neighbourhood of a particular vertex 
in $G$ and the robot progresses eastward from one column to the next
if and only if that robot has a transition corresponding to a vertex in the 
neighbourhood encoded in the first column. Subgrid SG2 thus checks if the 
robot encodes an actual dominating set of size $k$ in $G$, such that the robot 
enters the northeastmost square of the environment and builds the requested
structure there if and only if the $k$ east-moving transitions in the robot 
encode a dominating set of size $k$ in $G$.

Given the above, consider the following reduction from {\sc Dominating
set} to ContDesLS$_{res,D}$, Given an instance $\la G = (V,E), k\rangle$
of {\sc Dominating Set}, construct an instance $\langle E', E'_T, |T|, p_I, |L|,
r, |Q|, d, h, \#_{ec}\rangle$ of ContDesLS$_{res,D}$ as follows: Let $E'$ be 
the environment constructed in Lemma 5 of \cite{WV18_SI} with the
northwest $e_N$-based and SG1 subgrids removed, $E'_T$ be the version of $E_T$ 
in that same lemma, $p'_I = E'_{1,1}$, $L = \{\la q, enval(y, (0,0)), *, goEast, 
q'\ra ~ | ~ y \in \{e_1, \ldots, e_{|V|}\}\} \cup \{\la q, *, *, goNorth, q'\ra\}$, 
$|T| = |Q| = h = 1$, $r = 0$, $d = k + 1$,
and $\#_{ec} = 0$. Finally, let the target configuration of the task
associated with this instance be an FSR positioned in the northeastmost
square in $E'$; let us call this position $p_F$.
This instance of ContDesLS$_{res,D}$ can be constructed in 
time polynomial in the size of the given instance of {\sc Dominating Set}. 

Observe that the use of $L$ means that we no longer need subgrid SG1 and the 
restrictions on $|f|$ posited in Lemma 5 mentioned above to force the created
FSR to have $k$ east-moving transitions corresponding to a candidate dominating
set of $k$ distinct vertices in $G$. Hence, by slight simplifications and 
modifications of the proof of correctness of the reduction in Lemma 5 mentioned above,
it can be shown that there is a dominating set of size $k$ in graph $G$ in the 
given instance of {\sc Dominating set} if and only if there is an FSR with the 
structure specified in the constructed instance of ContDesLS$_{res,D}$ such 
(1) the lone FSR in $T$ can progress to $p_F$ if this FSR starts at 
$p_I$ and (2) the $k + 1$ transitions in this FSR are $k$ east-moving 
transitions from $L$ whose activation-formula predicates correspond to the 
vertices in a dominating set of size $k$ in $G$, and the final transition 
in $L$. As each transition in this FSR has an activation-formula consisting of 
either $*$ or a single predicate evaluating if that square has a particular 
square-type, there can be at most one transition enabled at a time and the 
operation of this FSR in $E'$ is deterministic. As the single robot in $T$ can 
only move north or east and does one of either in each move, the number of 
transitions executed in this construction task is the Manhattan distance from 
$p_I$ to $p_F$ in $E$. This distance is $(|V| + 1) + (|V|^2 + 1) < |E'| = 
c_1|E'|^{c_2} < c_1(|E'| + |Q|)^{c_2}$ when $c_1 = c_2 = 1$, which
means that this navigation task is $(1,1)$-completable.
As {\sc Dominating set} is $NP$-complete \cite[Problem GT2]{GJ79}, the reduction
above establishes that ContDesLS$_{res,D}$ is $NP$-hard; our main result then 
follows from Lemma \ref{LemAppProp1}. 

We now need to establish the $NP$-hardness of ContDesLS$_{res,D}$ for fixed values of 
$h > 1$, $|Q| > 1$, and $\#_{ec} > 0$. The case of $\#_{ec}$ can be 
handled by mechanisms and logic analogous to those used 
for ContEnvVer in the proof of Result A. The modifications
required in the case of $h$ and $|Q|$ are more complex. Let us first consider
the modifications required when $h = 1$ and $|Q| > 1$:

\begin{enumerate}
\item Add square-types $e_{q1}, e_{q2}, \ldots, e_{q(|Q| - 1)}, e_E$ to $E_T$.
\item Add transition-templates $\{\la q, enval(e_{qi},(0,0)),goSouth, q'\ra,
                         \la q, enval(e_{qi},(0,0)),$ \linebreak $goEast, q'\ra 
                        ~ | ~ 1 \leq i \leq |Q| - 1\} \cup
                       \{\la q, enval(e_E,(0,0)), goEast,q'\ra\}$ to $L$.
\item Create a subgrid area $SGA$ of $|Q|$ columns, where column $j$ includes a subcolumn 
       consisting of the square-types $e_{q1}, e_{q2}, \ldots, e_{qj}$ if 
       $1 \leq j \leq \min(k + 2, |Q| - 1)$ and
       a subcolumn consisting of the square-types $e_{q1}, e_{q2}, \ldots, e_{q(k + 2)},
       q_{ej}$ if $\min(k + 2, |Q| - 1) + 1 \leq j \leq |Q| - 1$. Let the bottom square
       of the subcolumn in column $j$ be immediately to the west of the top square of the subcolumn in
       column $j + 1$ for $1 \leq j \leq |Q| - 2$, and there be a single square of
       type $e_E$ immediately to the east of the bottom square in the subcolumn
       in column $|Q| - 1$,
\item Create a new environment $E''$ by positioning the subgrid $SGA$ specified in (3) in $E'$
       to the immediate east of SG2 such that the square of type $e_E$ is to the 
       immediate east of the southwestmost square of SG2. Fill all previously unspecified
       squares in $E''$ with the square-type $e_N$.
\item Reset $k'$ to $k + 2$.
\item Reset $p_I$ to the northwestmost square of $SGA$ in $E''$.
\end{enumerate}

\noindent
In order to progress eastward from column $j$ to $j + 1$, $1 \leq j \leq |Q| - 1$, the 
current state $q$ of the FSR must have the transition $\la q, enval(e_{qj},(0,0)), goEast,
q'\ra$. However, to progress southwards from the beginning of the subcolumn to the
bottom of the subcolumn in column $j$, the current state $q$ of the FSR must also
have the transitions in the set $\{\la q, enval(e_{qi},(0,0)), goSouth, q\ra ~ | ~ 
1 \leq i < \min(j - 1, k + 1)\}$. Recall that the rules of deterministic FSR
operation require that there cannot be more than one transition enabled at a
time. Hence, an FSR that can successfully navigate the first $|Q| - 1$ columns
of subgrid $SGA$ must have a different state for each column of $SGA$. To then enter and traverse SG2,
the state $q_{|Q|-1}$ of the FSR must have the transitions 
$\la q_{|Q| - 1}, enval(e_E,(0,0)), goEast, q_{|Q|-1}\ra$,
$\la q_{|Q| - 1}, *, goNorth, q_{|Q|-1}\ra$, and $k'' \leq k'$ eastward-moving
transitions that correspond to a dominating set of size $k'' \leq k$ in graph $G$
in the given in stance of {\sc Dominating set}. In the case when  $h > 1$, further
modify $E''$ to include an additional ``holding area'' consisting of $h - 1$ squares 
enclosed by obstacle-squares such that no FSR inside this area can leave it, and populate
this holding area with arbitrary FSR based on $|Q|$ states such that none of these FSR 
makes an environmental modification. As $h$ and $|Q|$ are of fixed value, the modified 
instance of ContDesLS$_{res,D}$ can still be constructed in time polynomial in the size 
of the given instance of {\sc Dominating set}. Moreover, the reduction from 
{\sc Dominating set} to this modified instance of ContDes:S$_{res,D}$ shows the 
$NP$-hardness of ContDesLS$_{res,D}$ for the specified values of $h$ and $|Q|$.
\end{proof}

\vspace*{0.1in}

Result C requires a very straightforward modification of the algorithm 
presented previously in \cite{War19} to incorporate our new task model.

\vspace*{0.1in}

\noindent
{\bf Result C} (Modified from Result A in \cite{War19}): 
                TeamDesLS$_{res}$ is exact polynomial-time tractable when $|Q| \geq 1$, 
                 $h = 1$, and $\#_{ec} \geq 0$.
 
\vspace*{0.1in}

\begin{proof}
The algorithm in the proof of Result A in \cite{War19}, which tests for each 
controller $c$  in $L$ whether a team based entirely on $c$ can
construct $X$ at $p_X$ in at most $c_1(|E| + |Q|)^{c_2}$ timesteps, operates in
polynomial time. We need only modify that algorithm such that after each
timestep of robot team operation we check if the target configuration associated with 
the task is in $E$, which can also be done in polynomial time.
\end{proof}

\vspace*{0.1in}

Part (1) of Result D requires relatively straightforward modification to
environment and FSR transitions to force $h$ to be exactly 2 (as the proof
presented previously in \cite{WdH+Sub} only required that $h \leq 2$).
Part (2) requires relatively straightforward modification
of the construction used in the proof of part (2) of Result A to allow $h > 2$ and
$|Q| > 1$.

\vspace*{0.1in}

\noindent
{\bf Result D} (Modified from Lemma A.7 in \cite{WdH+Sub}): 
                TeamDesLS$_{res}$ is not exact polynomial-time tractable when $|Q| = 1$, 
                 $h = 2$, and $\#_{ec} = 0$. Moreover, this
                 intractability holds for any version of TeamDesLS$_{res}$ when
                 $h > 2$, $|Q| \geq  1$, and $\#_{ec} \geq 0$.  
 
\vspace*{0.1in}

\begin{proof}
Lemma A.7 in \cite{WdH+Sub} gives a reduction from 3SAT to a version of problem SelAlg$_D$
which is for all intents and purposes our problem TeamDesLS$_{res,D}$ without the 
restriction on $\#_{ec}$. In the
instance of SelAlg$_D$ created by this reduction, a team $T$ of $|U|$ single-state FSR
chosen from an FSR library of size 2 encodes a truth assignment to the variables in
$U$ in the given instance of 3SAT and each of the southmost $|C|$ rows in environment $E$
encodes a clause in $C$ in this same instance. All robots in the team can reach
the northmost row of $E$ and deposit a strip of $|T|$ squares of type $e_X$
if and only the truth-assignment to the variables encoded in $T$ satisfies all clauses
in $C$; moreover, as at least one robot moves north in each timestep 
and strictly less than $|E|$ such moves can be made
if $T$ encodes such a valid truth-assignment, the task is $(1,1)$-completable.

We modify this reduction to create a reduction from 3SAT to TeamDesLS$_{res,D}$ when 
$h = 2$ and $\#_{ec} = 0$ as follows:

\begin{enumerate}
\item Add new clauses $c_T$ and $c_F$ to $C$ and new variables $u_T$ and $u_F$ to $U$ 
       such that clause $c_T$ ($c_F$) is satisfied if and only if variable $u_T$ ($u_F$) 
       is assigned value $True$ ($False$).
\item Add two extra columns to $E$ corresponding to new clauses $c_T$ and $c_F$ and
       add two to the size of the robot team.
\item In each of two robots in $L$, replace the transition that modifies the type
       $e_B$ of a square in the northmost row of $E$ to $e_X$ with a transition that
       stays on a square of type $e_B$.
\end{enumerate}

\noindent
Note that this reduction creates an FSR team in which $h = 2$, $|Q| = 1$, and
$\#_{ec} = 0$. If we then make the target configuration of the task associated
with this instance of TeamDesLS$_{res,D}$ be $|T|$ FSRs positioned in the
central squares of the northmost row of $E$, the proof of reduction
correctness in Lemma A.7 also establishes that this reduction is correct. As 
3SAT is $NP$-hard \cite[Problem L02]{GJ79}, this reduction establishes that 
TeamDesLS$_{res,D}$ is also $NP$-hard when $h = 1$, $|Q| = 1$, and $\#_{ec} = 0$;
our main result then follows from Lemma \ref{LemAppProp1}.

To complete the proof, observe that to establish the $NP$-hardness of \linebreak TeamDesLS$_{res,D}$
and polynomial-time unsolvability unless $P = NP$ of TeamDesLS$_{res}$ for fixed values of 
$h > 2$, $|Q| > 1$, and $\#_{ec} > 0$, we can use appropriately modified versions of the
constructions and logic that we used to show similar results for TeamEnvVer in the proof 
of Result A above.
\end{proof}

\vspace*{0.1in}

Finally, in Result E, inapproximability in senses (1) and (2) 
follows from the previously presented proofs in \cite{WdH+Sub} and 
the $NP$-hardness of decision versions of our verification and design 
problems shown in Results A, B, and D above. Inapproximability in sense (3)
is new to the current paper but follows in a very straightforward manner
from a known class inclusion result in computational complexity theory and 
the inapproximability proof for sense (2).

\vspace*{0.1in}

\noindent
{\bf Result E} (Modified from Results A.4 and A.5 in \cite{WdH+Sub}): 
                None of the intractable versions of TeamEnvVer, ContDesLS$_{res}$,
                 or TeamDesLS$_{res}$ described in Results A, B, and D are
                 polynomial-time approximable in senses (1--3) unless $P \neq BPP$ and 
                 $P = NP$.
 
\vspace*{0.1in}

\begin{proof}
That approximate polynomial-time solvability in sense (1) for any of the listed problems 
implies $P = NP$ follows from the $NP$-hardness of the decision versions of each of these
problems (which is established in the proofs of Result A, B, and D) and Corollary 2.2 in 
\cite{HW12}. 

With respect to approximate polynomial-time solvability in sense (2),
it is widely believed that $P = BPP$ \cite[Section 5.2]{Wig07} where $BPP$ is 
considered the most inclusive class of decision problems that can be efficiently
solved using probabilistic methods (in particular, methods whose probability of
correctness is $\geq 2/3$ and can thus be efficiently boosted to be arbitrarily
close to one). Hence, if any of the listed problems
has a probabilistic polynomial-time algorithm
which operates correctly with probability $\geq 2/3$ then the decision version of that
 problem is by
definition in $BPP$. However, if $BPP = P$ and we know that each of the decision versions
of the listed problems
is $NP$-hard by the proofs of Results A, B, and D, this would then imply by the 
definition of $NP$-hardness that $P = NP$.

With respect to approximate polynomial-time solvability in sense (3),
it is known that $ZPP \subseteq BPP$, where $ZPP$ is the class of decision problems
that can be always be solved correctly by algorithms with expected polynomial runtime
\cite{Gil77}. Hence, if any of the listed problems is approximately solvable in sense (3) 
then the decision version of that problem is by definition in $ZPP$ as well as $BPP$. 
However, as each of the decision versions of the listed problems
is $NP$-hard by the proofs of Results A, B, and D, this would then imply by the 
definition of $NP$-hardness and the widely-believed conjecture $P = BPP$ that $P = NP$.
\end{proof}

\end{document}